# Improving the accuracy of hard photon emission by Sigmoid sampling of the QED-table in particle-in-cell-Monte-Carlo simulations

Yinlong Guo[1,2], Xuesong Geng [1*], Liangliang Ji[1†], Baifei Shen[1,3] and Ruxin Li [1,4‡]

[1]State Key Laboratory of High Field Laser Physics and CAS Center for Excellence in Ultra-intense Laser Science, Shanghai Institute of Optics and Fine Mechanics (SIOM), Chinese Academy of Sciences (CAS), Shanghai, 201800, China.

[2] Center of Materials Science and Optoelectronics Engineering, University of Chinese Academy of Sciences, Beijing, 100049, China.

[3] Shanghai Normal University, Shanghai, 200234, China.

[4] Shanghai Tech University, Shanghai, 201210, China.

**Abstract**

Research on laser-plasma interaction in the quantum-electrodynamic (QED) regime has been greatly advanced by particle-in-cell & Monte-Carlo simulations (PIC-MC). While these simulations are widely used, we find that noticeable numerical error arises due to inappropriate implementation of the quantum process accounting for hard photon emission and pair production in the PIC-MC codes. The error stems from the low resolution of the QED table used to sample photon energy, which is generated in the logarithmic scale and cannot resolve high energy photons. We propose a new sampling method via Sigmoid function that handles both the low energy and high energy end of the photon emission spectrum. It guarantees the accuracy of PIC-MC algorithms for hard photon radiation and other related processes in the strong-field QED regime.

\* xsgeng@siom.ac.cn
† jill@siom.ac.cn
‡ ruxinli@mail.siom.ac.cn

# 1. Introduction

The 10-100 PW laser facilities around the world are committed to provide laser pulses of intensity approaching $10^{23}$W/cm$^2$ [1–3]. Under such condition laser-matter interaction exhibits new features, such as emission of high energy gamma photons and production of electron-positron pairs. These important processes can be described by the strong-field quantum electro-dynamics (SF-QED) theory [4]. Due to the high nonlinearity in relativistic laser-plasma and SF-QED regime, it is unlikely to obtain accurate analytical solutions for the interactions. Therefore, computational methods based on the particle-in-cell (PIC) algorithm [5] and Monte-Carlo (MC) sampling have been developed to simulate light-matter interactions in this regime [6–10].

In open-sourced codes like EPOCH and SMILEI [11,12], the hard photon emission and pair production are implemented into regular PIC loop via the MC approach [13–16] to capture the probabilistic and stochastic QED effects [14,17–19]. To optimize the computation efficiency, a pre-calculated table is set from the SF-QED theory [20], according to which stochastic QED events are triggered after MC sampling and fine interpolation. Here we use the code SMILEI to simulate the collision between an ultrashort high intensity laser pulse and an ultra-relativistic electron bunch. We find unusual stair-like structure in the high energy end of the photon spectra in moderate QED regime. We point out that the occurrence of the structure originates from insufficient/ inappropriate QED table sampling. This leads to significant overestimation of photon emission near the spectrum cut-off. The error becomes more apparent for larger QED parameters, e.g., higher electron energy, stronger laser fields and/or shorter pulse duration. The disparity can be slightly mitigated by using large table size for larger QED parameters, which, however, cannot solve the problem once and for all. In this work, we propose a new sampling algorithm that can well handle the sampling of the photon emission spectrum in both the low-energy end and the high-energy end at reasonable table size. The new method safely extends the application of QED-MC algorithm to larger QED parameters and more extreme situations.

## 2. Method

### 2.1 Theoretical basis

The regular loop of PIC algorithm is modified to account for nonlinear Compton scattering and nonlinear Breit-Wheeler pair production [11,12]. In laser-plasma interactions, charged particles follow the Lorentz equation $\frac{dp}{dt} = e(\boldsymbol{E} + \boldsymbol{v} \times \boldsymbol{B})$ when no QED processes are triggered. Here $\boldsymbol{p} = \gamma m_e \boldsymbol{v}$ is the particle momentum, with $m_e$ the particle rest mass, $e$ the particle charge, $\boldsymbol{E}$ and $\boldsymbol{B}$ the electric and magnetic field, respectively. Electrons moving in laser-plasma fields radiate, which is defined by the nonlinear QED parameter $\chi_e \sim \gamma \frac{|E|}{E_{cr}}$, where $\gamma$ is the electron relativistic factor and $\boldsymbol{E}_{cr} = \frac{(m_e^2 c^3)}{\hbar e} \approx 1.3 \times 10^{16}$ V/cm the Sauter-Schwinger critical field [21]. The quantum effect and strong radiation-reaction becomes active when $\chi_e$ approaches unity at large normalized laser amplitude $a_0 = \frac{eE}{m_e \omega_0 c} \gg 1$ ($\omega_0$ is the laser angular frequency) since $\chi_e$ controls the strength of nonlinear quantum effects [22–24].

Considering ultra-relativistic electrons, the emitted photons gather in a radiation cone $\Omega \sim \frac{1}{\gamma} \ll 1$, almost parallel to the electron momentum. For strong laser field ($a_0 \gg 1$) the coherence length is $\sim \frac{\lambda}{a_0} \ll 1$, i.e., the scale of the electromagnetic field variation is quite large as compared to the radiation formation length, allowing one to employ locally-constant cross-field approximation (LCFA) [25] at field strengths much smaller than the Schwinger limit $\frac{E}{E_{cr}} \ll 1$, which is proved to be numerically valid when fields are sufficiently strong [26,27]. Under these assumptions, the point-like QED processes take place on the classical trajectory. The Lorentz invariant rate of

instantaneous photon emission of an electron is [18,22]:

$$\frac{d^2 N_\gamma}{d\tau d\chi_\gamma} = \frac{2}{3}\frac{\alpha^2}{\tau_e}\frac{G(\chi_e,\chi_\gamma)}{\chi_\gamma}, \tag{1}$$

where $\alpha \approx 1/137$ is the fine structure constant, $\tau_e = \frac{r_e}{c} = \frac{e^2}{4\pi\epsilon_0 m_e c^3}$ the time for light to cross the classical radius of the electron, $G(\chi_e,\chi_\gamma) = \frac{\sqrt{3}}{2\pi}\frac{\chi_\gamma}{\chi_e}[\int_v^{+\infty} K_{\frac{5}{3}}(y)dy + \frac{3}{2}\chi_\gamma v K_{\frac{2}{3}}(v)]$, $K_{\frac{1}{3}}(y)$, $K_{\frac{2}{3}}(v)$ are modified Bessel functions and $v = \frac{2\chi_\gamma}{3\chi_e(\chi_e-\chi_\gamma)}$, $y = \frac{\chi_\gamma}{3\chi_e(\chi_e-\chi_\gamma)}$, respectively. The Lorentz-invariant quantum parameter is composed of the electron four-momentum $p_\nu$ ( a photon with four-momentum $\hbar k_\nu$) and field tensor $F^{\mu\nu}$ [4,22,23]:

$$\chi_e = \left|\frac{F^{\mu\nu}}{E_s}\frac{p_\nu}{m_e c}\right|, \text{ and } \chi_\gamma = \left|\frac{F^{\mu\nu}}{E_s}\frac{\hbar k_\nu}{m_e c}\right|. \tag{2}$$

The parameter $\chi_e$ decides the character of photon emission: for classical radiation $\chi_e \ll 1$, there can be large numbers of photons and each one contains very small portion of energy of the emitting electron. In consequence, the radiation reaction could be modelled classically as a continuously damping force. In moderate quantum region $\chi_e \sim 1$, during each emission the photon may take a very large part of the electron energy thus hard photons are emitted discreetly and stochastic quantum effects arise [18,23].

High energy photons can further decay into $e^+e^-$ pairs in strong fields, known as the nonlinear Breit-Wheeler process, dominated by the parameter $\chi_\gamma$. Similarly, the Lorentz invariant rate of photon decaying into $e^+e^-$ pairs is [18,22]:

$$\frac{d^2 N_\pm}{d\tau d\chi_-} = \frac{1}{\sqrt{3}\pi}\frac{\alpha m_e c^2 \chi_e}{\hbar \chi_\gamma^2}\int_v^{+\infty}\sqrt{y}K_{\frac{1}{3}}\left(\frac{2}{3}y^{\frac{3}{2}}\right)dy - (2-\chi_\gamma x^{\frac{3}{2}})K_{\frac{2}{3}}\left(\frac{2}{3}x^{\frac{3}{2}}\right), \tag{3}$$

where $\chi_-$ is quantum parameter of the newborn electron, and $x = \left(\frac{\chi_\gamma}{\chi_e} + \chi_-\right)^{\frac{2}{3}}$.

## 2.2 Implementation of QED modules in PIC

The numerical models of QED in PIC vary in different codes. But they share a common idea of implementation [11,12,14–16,19]. We introduce the MC realization of photon emission and pair production in SMILEI [12] based on the theoretical approach introduced above. For photon emission, every propagating electron in the laser field is assigned with an optical depth $\tau_e$ with initial $\tau_{e0} = 0$. Its time evolution follows [12,18]:

$$\frac{d\tau_e}{dt} = \int_0^{\chi_e} \frac{d^2 N_\gamma}{d\tau d\chi_\gamma} d\chi_\gamma. \tag{4}$$

When the optical depth $\tau_e$ reaches (larger than) the optical depth $\tau_f$, a photon is emitted along the direction of electron motion. Since the photon radiation event is similar to the particle decay, which follows Poisson distribution $e^{-\lambda}$, the optical depth $\tau_f$ can therefore be randomly sampled determined by $\tau_f = -\ln(\xi)$ where $\xi \in (0,1)$ is a uniform random number.

Once the photon radiation is triggered by reaching $\tau_f$, the photon energy is randomly chosen from the radiation spectrum $\frac{d^2 N_\gamma}{d\tau d\chi_\gamma}$, as shown in Fig. 1(a). Here the radiation spectrum becomes the probability density function (PDF) of the energy that defines the weight of different photon energies when randomly choosing a photon energy. Therefore, by integrating the PDF along the photon energy axis one gets the sampling function [12,18]:

$$P_{\chi_e}(\chi_\gamma) = \frac{\int_0^{\chi_\gamma} \frac{d^2 N_\gamma}{d\tau d\chi_\gamma} d\chi_\gamma}{\int_0^{\chi_e} \frac{d^2 N_\gamma}{d\tau d\chi_\gamma} d\chi_\gamma} \tag{5}$$

that stacks the weights of different energies, where the denominator scales the function to $(0, 1)$, as shown in Fig. 1(b). Then uniform random sampling of $P_{\chi_e}(\chi_\gamma)$ can generate photon energies of which the distribution is identical to the PDF. This process requires calculating the inverse function of $P_{\chi_e}(\chi_\gamma)$ and the photon energy can then be calculated by:

$$\chi_\gamma = P_{\chi_e}^{-1}(\xi') \tag{6}$$

where $\xi' \in (0,1)$ is another random number, as shown by the arrows in Fig. 1(b). For example, when $\xi' = 0.97$ and $\chi_e = 1$, it reads $\delta \approx 0.58$, as shown by the arrows in Fig. 1(b). Once $\chi_\gamma$ is determined, the photon energy is $\epsilon_\gamma = \gamma_e m_e c^2 \frac{\chi_\gamma}{\chi_e}$. Through momentum conservation, i.e., $p_e^f =$

$\left(p_e^i - \frac{\epsilon_\gamma}{c}\right) = (1-\delta)p_e^i$, discrete and stochastic photon emission and radiation reaction are numerically implemented. We should note that the $\chi_\gamma$ axis is normalized to $\delta = \frac{\chi_\gamma}{\chi_e} \in (0,1)$ in the calculations.

Using the above sampling method, we generate $10^6$ photons from $10^6$ random numbers $\xi'$ at specific $\chi_e$s, and their spectrums are in well agreement with the analytical spectrum $\frac{d^2N_\gamma}{d\tau d\chi_\gamma}$ for $\chi_e = 0.01, 1, 10$, as shown in Fig. 1(a). For computational accuracy, the evaluation of $P_{\chi_e}^{-1}$ uses a look-up table of size $N = 1024$ (see Section 2.3 for explanation).

As for BW pair production, the MC scheme follows a similar way, with the rate defined by Eq. (3). Note that only photons with energy higher than $2m_ec^2$ are involved in pair production, accounting for the rest mass of the generated pair.

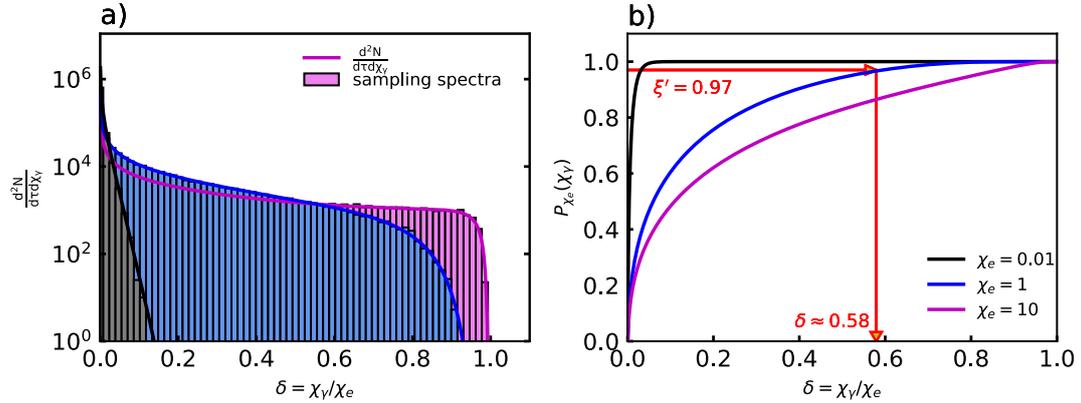

**Figure 1.** (a) The normalized photon emission rates $\frac{d^2N_\gamma}{d\tau d\chi_\gamma}$ (dotted-dashed lines) and the spectrums of $10^6$ photons from random sampling (bars). (b) the cumulative integrals of spectrums in (a). The red arrows indicate the interpolation of $\delta_\gamma = \chi_\gamma/\chi_e$ from $\xi_\gamma$. The results shown in black, blue and purple correspond to $\chi_e = 0.01, 1$ and $10$.

## 2.3 Tabulation and interpolation of the QED emission probability

The QED process in PIC implementation is computationally expensive as compared to the PIC

loop due to the complexity of the integral functions. To improve the simulation efficiency of the QED process, the integrals are approximated by interpolation from pre-calculated tables sampled at pre-defined values, e.g., $\chi_e$ and photon energies. This technique is widely accepted by PIC simulation codes [11,12].

For computation of the optical depth $\tau_e$, tabulation of Eq. (4) at different $\chi_e$ is required. To determine the generated photon energy, Eq. (5) is further tabulated for different $\chi_e$ and $\chi_\gamma$. In both equations, $\chi_e$ is usually discretized in logarithmic scale in order to cover a wide range of interaction, e.g., $\chi_e \in (10^{-4}, 10^3)$. Then Eq. (4) is integrated along $\chi_\gamma$ and tabulated with chosen $\chi_e$ values. In Eq. (5) the photon energy parameter $\chi_\gamma \in (0, \chi_e)$ is also discretized in logarithmic scale in general cases for fine sampling of the low energy photons since the radiation probability of high energy photons ($\chi_\gamma \approx \chi_e$) is low.

At each time step, $\log_{10}(\chi_e)$ is calculated to determine the location for interpolation in the table of Eq. (4). Then $\frac{d\tau_e}{dt}$ is interpolated from the data near $\log_{10}(\chi_e)$, e.g., $\left(\frac{d\tau_e}{dt}\right)^{(m)}$ and $\left(\frac{d\tau_e}{dt}\right)^{(m+1)}$. Linear interpolation is widely adopted in this case such that $\frac{d\tau_e}{dt}$ is determined by

$$\frac{d\tau_e}{dt} = \frac{\left(\frac{d\tau_e}{dt}\right)^{(m+1)} - \left(\frac{d\tau_e}{dt}\right)^{(m)}}{\chi_e^{(m+1)} - \chi_e^{(m)}} \times \left[\chi_e - \chi_e^{(m)}\right] + \left(\frac{d\tau_e}{dt}\right)^{(m)}, \quad (7)$$

where $\chi_e^{(m)}$ is logarithmically distributed. If a photon emission event is triggered, the photon energy is determined by inverse sampling of Eq. (5) from a uniformly distributed random number $\xi \in (0, 1)$, i.e., $\chi_\gamma = P_{\chi_e}^{-1}(\xi)$, as shown in Fig. 1 (b). Photon energy parameter $\chi_\gamma$ is interpolated by inversing the curve defined in the table of Eq. (5) at a given $\chi_e$

$$\chi_\gamma = \frac{\chi_\gamma^{(m+1)} - \chi_\gamma^{(m)}}{P_{\chi_e}^{(m+1)} - P_{\chi_e}^{(m)}} \times \left[\xi - P_{\chi_e}^{(m)}\right] + \chi_\gamma^{(m)}. \quad (8)$$

Again, $\chi_\gamma \in (0, \chi_e)$ is also logarithmically distributed. It should be noted that the radiation probability in Eq. (1) diverges at low photon energies thus the lower limit of $\chi_\gamma$ in the integration of Eq. (5) must be manually chosen [12,15].

The BW process shares a similar way for tabulation and random sampling defined by Eq. (3) and its integrals, except that the BW spectrum peaks at $\delta_\pm = \frac{\chi_\pm}{\chi_\gamma} = 0.5$ and is symmetric to $\delta_\pm = 0.5$, meaning that it only requires tabulation of $\delta_\pm \in (0, 0.5)$.

## 3. Results

### 3.1 Error in photon spectra in the QED regime

In our 2-dimensional simulations, the laser pulse of pulse duration $\tau_L$ (full width at half maximum, FWHM) and normalized amplitude $a_0$ head-on collides with an electron bunch of kinetic energy $\varepsilon_e = \gamma m_e v^2$. In other to emphasize the numerical errors of the Monte-Carlo method, mono-energetic electron bunch is utilized for demonstration. The wavelength of the laser pulse is $\lambda_L = 0.8$ μm and the focal spot size is $w_0 = 5$ μm $\approx 6.25\lambda_L$. This linearly polarized laser pulse takes the Gaussian profile $a_L = a_0 e^{-\frac{t^2}{\tau_L}} e^{-\frac{r^2}{w_0}} \sin(\omega_0 t) \boldsymbol{e_y}$ with $r^2 = y^2 + z^2$ ($z = 0$ for 2D simulation), $\omega_0 = \frac{2\pi c}{\lambda}$ and enters from the left of the simulation window. The electron bunch is also Gaussian profiled, $n_e(r, \xi = x - ct) = n_0 exp(-\frac{r^2}{2\sigma_r^2} - \frac{\xi^2}{2\sigma_x^2})$. Here $\sigma_r^2 = 0.5 \times 0.5$ μm$^2$, $\sigma_x^2 = 6.25 \times 6.25$ μm$^2$ the longitudinal and transverse beam sizes and $n_0 = 10^{-3} n_c$, with $n_c = \frac{m_e \omega^2}{4\pi e^2}$, where $e$ and $m_e$ the electron charge and mass, $\omega_0$ the field frequency. The grid size of the simulation is $\frac{\lambda_L}{160} \times \frac{\lambda_L}{40}$ in x- and y-direction to guarantee the accuracy of the simulation and the time step is $\frac{T_0}{100}$ with $T_0 = \frac{\lambda_L}{c}$. The length of the simulation window in x-direction ranges from 8 μm to 90 μm corresponding to pulse lengths from 2.5 fs to 100 fs and the length of y-direction is 20 μm.

The spectra of photons and electrons after collision are summarized in Fig. 2 for $a_0 = 100$ and $\varepsilon_e = 1000$ MeV (corresponding to $\chi_{e,max} \sim 1$). Here only the pulse duration $\tau_L$ changes, from sub-cycle (2.5 fs) to long pulses (100 fs).

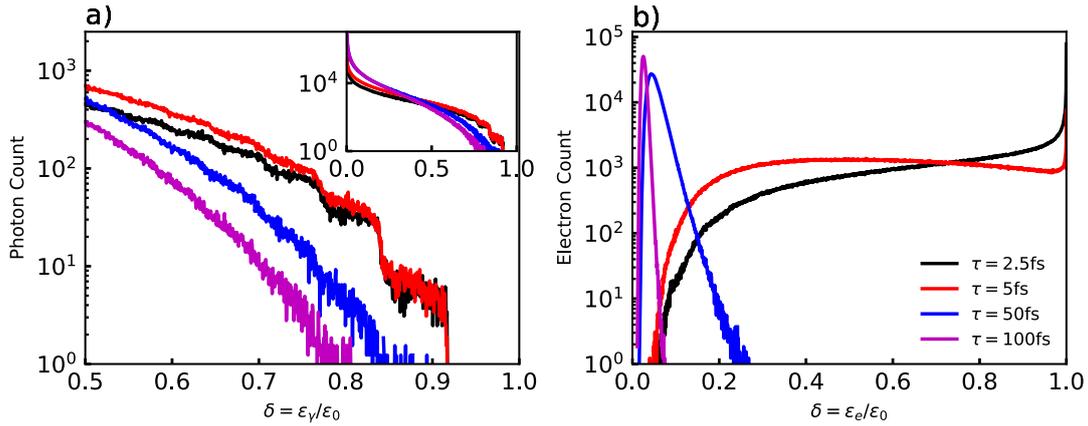

**Figure 2.** Energy spectra for photons (a) and electrons (b) after the interaction with $a_0 = 100$ and electron energy $\varepsilon_0$=1000 MeV. Laser pulse duration $\tau_L$ varies from 2.5 to 100 fs. Here $\delta = \frac{\varepsilon_{e,\gamma}}{\epsilon_0}$ and the inset in (a) shows the full energy spectra.

It is shown in Fig. 2 (a) that the photon cut-off energy increases for shorter pulses and longer pulses generate more low-energy photons. This is the stochastic nature of QED radiation: electrons radiate photons with certain probabilities. For shorter laser pulses, some electrons may undergo limited or even no photon emission events in the rising edge of the pulse. They reserve most of the energy when reaching the pulse center such that the electron QED parameter can be maximized. Photons emitted here carry largest amount of electron energies. On the contrary, longer pulses have more gentle pulse front, which depletes the electron energy by inducing more photon emission events. Before approaching the pulse center, electrons are already cooled down and the peak QED parameter is lower. As verified in Fig. 2 (b), there are plenty of electrons reserve their energies for few-cycle lasers but gathering in the low energy region for longer pulses.

It is noticed that stair-like structure appears at the high energy end of photon spectra for ultrashort pulse duration. The featured structure emerges while stochastic quantum effects become important. However, as we will show below, such structure reflects inaccurate sampling in the

Monte-Carlo process accounting for photon emission. This may cause huge error in the high energy end and incorrectly enhance the stochastic behavior at short pulse lengths [8].

### 3.2 The stair structure dependence on $a_0$ and $\varepsilon_e$

In the following simulations we choose three laser durations $\tau_L = 2.5\text{ fs}, 30\text{ fs}$ and $100\text{ fs}$. First, we fix the electron energy at $\varepsilon_e = 1000$ MeV and change the normalized laser amplitude from $a_0 = 20$ to $a_0 = 200$. Then the electron energy $\varepsilon_e$ varies from 100 MeV to 2000 MeV keeping the same laser amplitude $a_0 = 100$. The corresponding peak quantum parameter is $\chi \sim 0.1$, 1 and 2. Fig. 3 shows the spectra of photons after interaction. The stair structure appears when the cut-off of photon energy ratio $\delta = \varepsilon_\gamma/\varepsilon_0$ approaches unity. This is most evident for ultrashort pulse duration, large laser amplitude and high electron energies, as seen in Fig. 3 (a) and (d). Longer pulses employed in Fig. 3 (b) and (c) produce much smoother spectrum distributions, hence lower electron energy. Thus, the photon energy ratio $\delta \sim 1$ or QED parameter $\chi \sim 1$ is related to whether the stair structure appears or not.

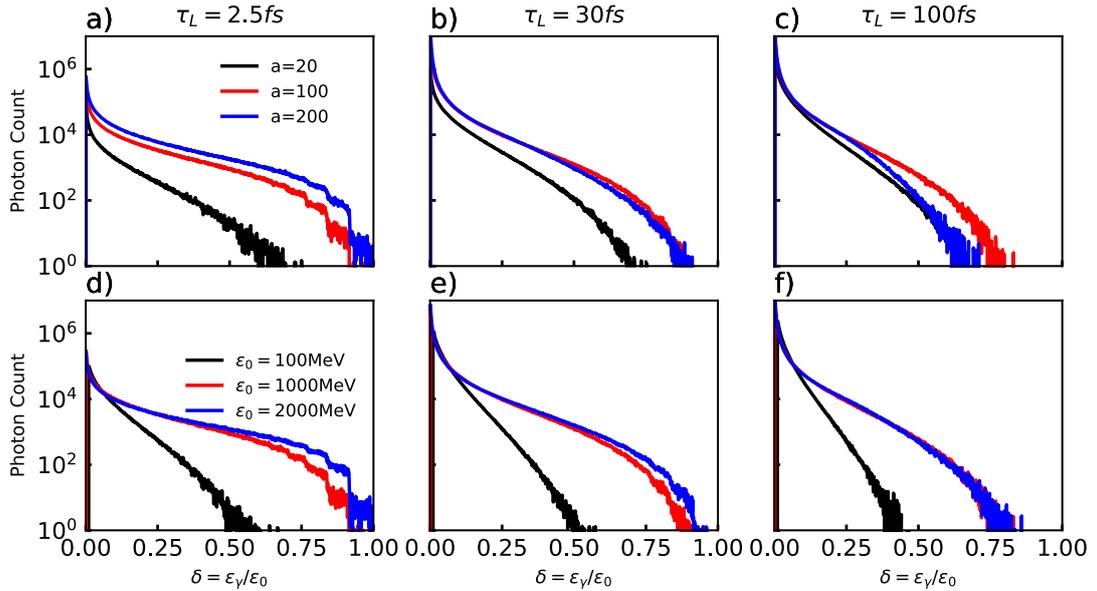

**Figure 3.** The photons energy spectra with different $\tau_L$ varies from 2.5 to 100 fs and $a_0$ varies from

20 to 200 (a-c), the electron energy $\varepsilon_e$ varies from 100 MeV to 1000 MeV (d-f), with $\delta = \frac{\varepsilon_{e,\gamma}}{\epsilon_0}$.

### 3.3 The stair structure with different table size

We illustrate the consequences of inaccurate table sampling using the following parameters: $a_0 = 100, \tau_L = 5$ fs, $\varepsilon_e = 3200$ MeV ($\chi_e \sim 3.2$, moderate quantum region). The grid number of the QED-MC table can be set as 128, 256, 512 and 1024, respectively. In Fig. 4 (a) and (b) we see that in the low energy region $\delta < 0.8$ all distributions are consistent. Significant divergence is seen in the region beyond. For 128 points grids, there is a sharp decline around $\delta = 0.8$, followed by slow and long slop extending to $\delta = 1$. Here the photon number is still at an unexpected high level. The quasi-plateau starts at $\delta = 0.9$ for 256 grids and the photon number at maximum energy is suppressed by several times. The distribution tends to converge for 512 and 1024 points grids.

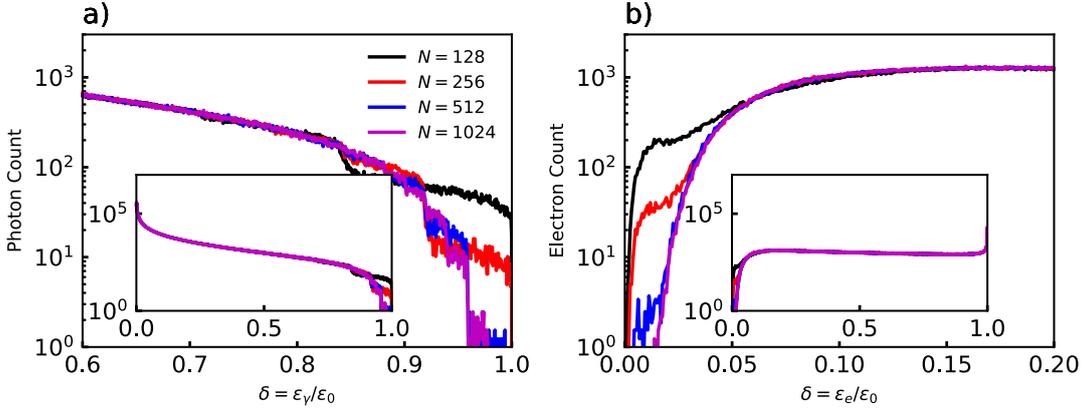

**Figure 4.** The energy spectra of photons (a) and electrons (b) with table size $N$= 128, 256, 512 and 1024. All the spectra are obtained at the same parameter $a_0 = 100, \tau_L = 5$ fs and $\varepsilon_e = 3200$ MeV.

Higher resolution apparently gives more precise results. The error caused by insufficient sampling resolution of the QED table overestimates the photon number emitted in the high energy end. Although these photons only cover a small region of the energy spectra, they exhibit strong signature of radiation-dominated QED effects, e.g., the number the low-energy electrons. As shown in Fig.

4(b), the low-energy electrons are also overcounted by 100 times at low resolution. Besides, the high-energy photons may decay into pairs through BW process in the intense pulse, further enlarge the error induced by over-estimated high-energy photons.

## 4. Discussion and Solution

### 4.1 Grid effect induced by logarithmic sampling

It has been shown in section 2.3 that the $\chi_\gamma/\chi_e$ values are logarithmically sampled for fine sampling of the low energy photons, resulting in coarse sampling of high energy photons. Fig. 5 (a-d) display the photon emission spectrum $\frac{d^2 N_\gamma}{d\tau d\chi_\gamma}$ for different $\chi_e$ and different table sizes $N$ and one can see sharp transitions of the spectrum when $\delta = \frac{\chi_\gamma}{\chi_e} > 0.5$, which is more significant when $\delta$ approaches unity. The grid effect can be eased by using larger table sizes but cannot be eliminated since the spacing between the gridded $\delta$ does not converge when $\delta \to 1$ under logarithmic sampling.

For chosen parameters $\chi_e = 0.01, 1, 10$, the spectra are compared in Fig. 6 (a-d), which can be seen as the featured radiation spectra under these parameters [19,22]. At small $\chi_e$, where the radiation can be treated classically, the spectra peak at low $\delta$ values. A few points are enough to rebuild the distribution. When $\chi_e$ is larger, the spectrum is significantly blue shifted and broadened. The peak even shifts towards $\delta \sim 1$ at $\chi_e = 10$. In this case, the logarithmically sampled photon energies cannot resolve the spectra. The change at the high energy end is so steep that polynomial fitting cannot approximate the spectrum, leading to the unphysical stair structure of the energy spectra in QED-PIC simulation. This grid effect will imprint these errors into the integrated probability rate $\int_0^{\chi_e} \frac{d^2 N_\gamma}{d\tau d\chi_\gamma} d\chi_\gamma$, which modifies the random sampling of photon energies as shown in the results.

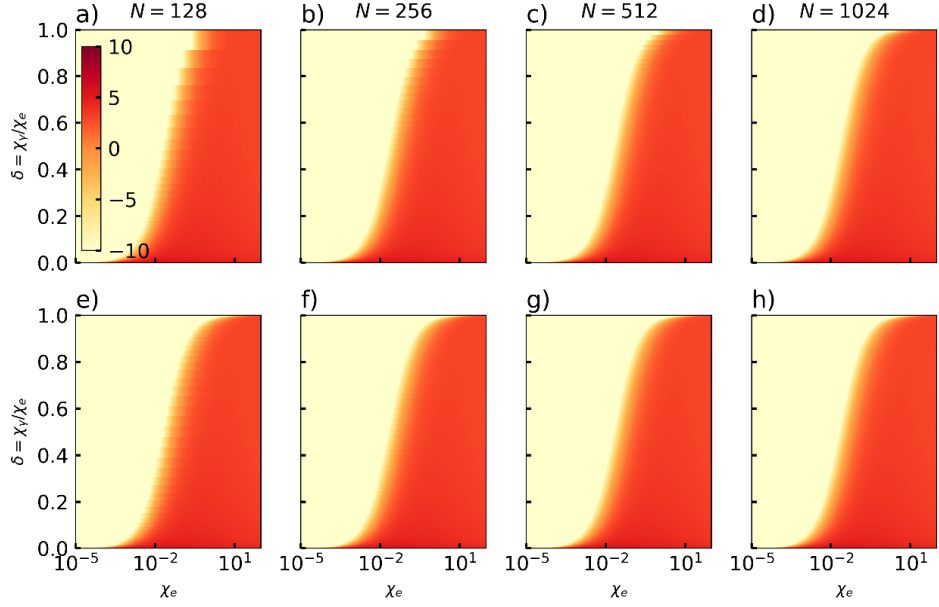

**Figure 5.** The photon emission spectrum $\log_{10} \frac{d^2 N_\gamma}{d\tau d\chi_\gamma}$, where $\delta = \frac{\chi_\gamma}{\chi_e} \in [10^{-5}, 1]$ is logarithmically sampled in (a)-(d) for table sizes of $N$=128, 256, 512 and 1024; (e)-(f) are the results of the modified Sigmoid sampling, where the parameter for Eq (11) is $A = 11.5$ thus $\delta \in [10^{-5}, 1 - 10^{-5}]$.

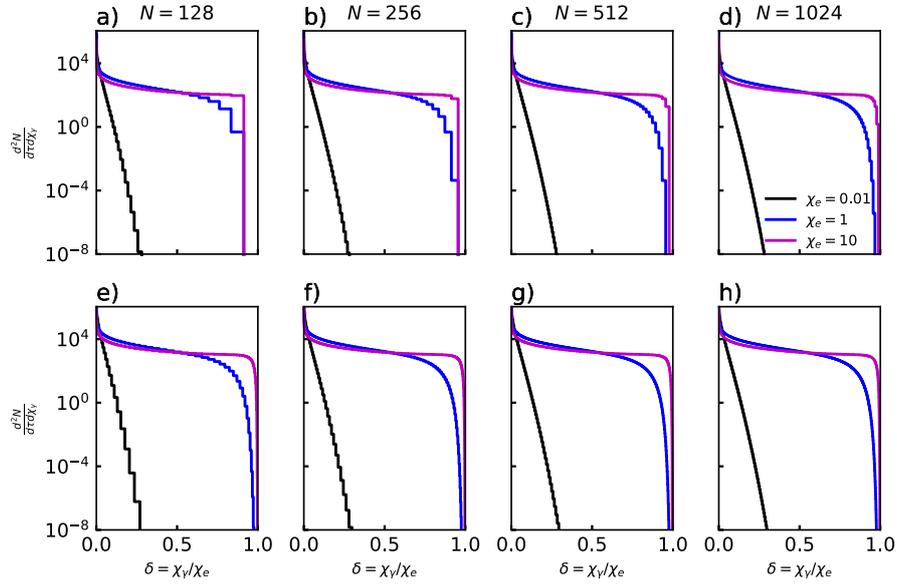

**Figure 6.** The photon emission spectrum $\frac{d^2 N_\gamma}{d\tau d\chi_\gamma}$ at $\chi_e = 0.01, 1, 10$: (a)-(d) represent the standard logarithmic sampling with $N$=128, 256, 512 and 1024, where $\delta = \frac{\chi_\gamma}{\chi_e} \in [10^{-5}, 1]$; (e)-(f) the data are calculated via Sigmoid function, where the parameter for Eq (11) is $A = 11.5$ thus $\delta \in$

$[10^{-5}, 1-10^{-5}]$.

The above results suggest that one must be careful when performing simulations including hard photon emission and pair production based on the QED table approach. According to the QED-MC implementation, it is possible to find an appropriate table size under certain conditions. For example, the maximum $\chi_e$ in laser-electron collision with laser amplitude $a_0$ and electron gamma factor $\gamma_i$ can be estimated by evaluating the electron energy in the laser pulse [23]:

$$\gamma_\tau = \frac{\gamma_i}{1+fa_0^2\gamma_i}, \tag{9}$$

where $f = \left(\frac{4r_e}{3}\right)\int_0^\tau \left[\frac{a_y(r,\tau')}{a_0}\right]^2 d\tau$ is the pulse profile factor and $\tau = \omega(t-\frac{x}{c})$ the relative phase. When the electron reaches the peak of the laser pulse, the peak QED parameter can be estimated by [30]:

$$\chi_{e,\max} = \frac{a_0\gamma_\tau\omega(1+\cos\theta)}{m} = \frac{2\omega}{m}\frac{a_0\gamma_i}{1+fa_0^2\gamma_i}. \tag{10}$$

For example, if $\chi_{e,\max} = 1$, the largest photon energy to resolve is $\delta_{\max} \approx 0.95$, as shown in Fig. 6. Supposing a table of size $N$ is discretized in logarithmic scale in the $\delta$ direction, i.e., $\delta \in [10^{-\alpha}, 10^0]$ and $\delta^{(m)} = 10^{-\alpha+m\frac{\alpha}{N-1}}$. If the acceptable resolution of $\delta$ near $\delta_{\max}$ is $\epsilon = \delta^{(m)} - \delta^{(m-1)} < 10^{-2}$, the minimum $N$ required is $N_{\min} = 1100$ for $\alpha = 5$, which can be varied depending on $\epsilon$.

As for BW process, $\delta_\pm$ is also logarithmically discretized but in the range of $[10^{-\alpha}, 10^{\log_{10} 0.5}]$ due to symmetry of the electron/positron spectrum as shown in Fig. 7(a). Since the slope of the spectrum at $\delta_\pm = 0.5$ is zero, the symmetric logarithmical discretization is sufficient to resolve the spectrum at $\delta_\pm = 0.5$ and $\delta_\pm \approx 0$ at the table size of $N = 128$, as shown in Fig. 7(b) where 255 data points are displayed due to symmetry. It should be noted that the SMILEI code implicitly avoided the grid effect via symmetric logarithmical discretization and will not induce significant numerical error like the photon spectrums.

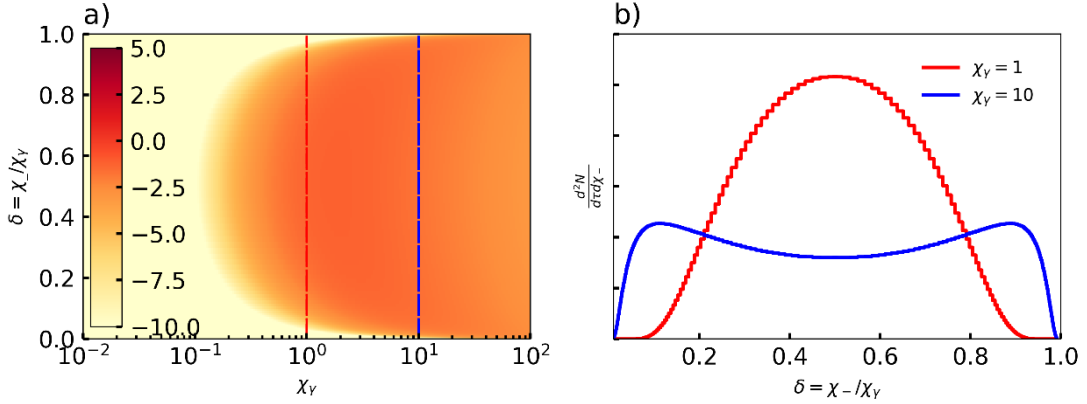

**Figure 7.** (a) The electron/positron spectrums of BW process. The dashed lines correspond to the spectrums in (b) the BW pair production spectra at $\chi_\gamma=1, 10$ with logarithmic discretization of $\delta \in (10^{-\alpha}, 0.5)$, the other side is drawn by symmetry.

**4.2 Sigmoid sampling**

Increasing the table size can ease the tabulation effects but logarithmic sampling can never mimic the asymptotic behavior in both low energy and high energy ends of the spectrum. Note that even if high-order polynomial fitting is used in the interpolation, the rapid change of $\frac{d^2 N_\gamma}{d\tau d\chi_\gamma}$ at strong QED region $\chi_e > 1$ ($\delta \sim 1$) can hardly be reproduced accurately. As a result, we encounter the stair-like energy spectra with incorrect energies and amounts for hard photons.

We notice the asymptotic behavior of the photon emissivity for high energy and low energy photons in Fig. 5, the latter of which was solved by using logarithmic distribution of $\delta = \chi_\gamma/\chi_e$ which results in the problem discussed here. To resolve both the high energy and low energy ends, we propose distributing the $\delta$ values of the QED table via Sigmoid function:

$$\delta = \frac{1}{1+\exp(-AX)} \in [\delta_{\min}, \delta_{\max}] \qquad (11)$$

where $X$ is a uniform sequence of length $N$ and $A$ is the scaling parameter controlling the bounds of $\delta$. For instance, for $X \in [-1,1]$, Fig. 8 shows the distribution for $A = 6.91, 11.5, 16.1$ corresponding to $\delta \in [10^{-3}, 1-10^{-3}]$, $\delta \in [10^{-5}, 1-10^{-5}]$ and $\delta \in [10^{-7}, 1-10^{-7}]$, where the sequence length is restricted to $N = 128$. By controlling $A$ and $N$, one is able to control the bounds and shape of the distribution of $\delta$. For results shown in Fig. 5 (e-h) and Fig. 6 (e-h), we choose A=11.5 ($\delta_{min} = 10^{-5}$) for comparison with logarithmic sampling of $\alpha = 5$. For common cases $A = 8.805$, corresponding to $\delta \in [1.5 \times 10^{-4}, 1 - 1.5 \times 10^{-4}]$, is a good choice that the variation at $\delta = 0.5$ is not too steep, which is adopted in the simulations in Fig. 9.

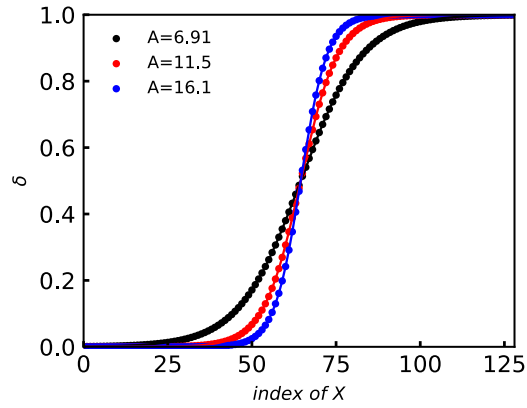

**Figure 8.** The shape and bounds of the Sigmoid sampled distribution for $A = 6.91, 11.5, 16.1$ corresponding to $\delta \in [10^{-3}, 1-10^{-3}]$, $\delta \in [10^{-5}, 1-10^{-5}]$ and $\delta \in [10^{-7}, 1-10^{-7}]$, where $X \in [-1, 1]$ is a uniform sequence of length $N = 128$.

Usually, the QED parameter $\chi_e$ ranges over deep quantum region with $\chi_{e_{max}} \gg 10^1$ and classical domain with $\chi_{e_{min}} \leq 10^{-3}$ for tabulation. Besides, the photon emission rate reaches a plateau after reaching quantum region as shown in Fig. 5 where the variation of the emission spectrum slows down for larger $\chi_e$. They can both help the accuracy in moderate quantum region $\chi_e \leq 10$. For these reasons, it is credible that suitable logarithmically sampled points $\chi_e \in (10^{-5}, 10^2)$ is enough for tabulation. One could just choose a wider range of $\chi_e$ to recede the numerical error for more extreme condition. Finally, for the tabulation of the integral of $\frac{d^2 N_\gamma}{d\tau d\chi_\gamma}$, $\delta =$

$\chi_\gamma/\chi_e$ is distributed via Sigmoid sampling and $\chi_e$ is logarithmically distributed.

**4.3 The comparison between two tabulation methods**

We compare the results of the proposed method with the logarithmic one in Fig. 5 (e-f). It is clear that even 128 points tabulating reproduces the characteristics of $\frac{d^2N_\gamma}{d\tau d\chi_\gamma}$ very well at the high nonlinear region $\delta \sim 1$, as well as the classical region $\delta \ll 1$, and better for $N$ = 256, 512 and 1024 points. Further, the newly discrete photon emissivity $\frac{d^2N_\gamma}{d\tau d\chi_\gamma}$ at chosen sections $\chi_e = 0.01, 1, 10$ are shown in Fig. 6 (e)-(f), where the stair structure is rarely noticeable, illustrating the high accuracy in all cases.

It should be noted that, as discussed in section 4.1, when using the Sigmoid sampling the required minimum table size to resolve $\epsilon = 10^{-2}$ near $\delta = 0.95$ for $\chi_e = 1$ is $N_{\min} = 60$, much lower than that of the logarithmic sampling. With a table of fair and appropriate size it can be helpful for efficient and economical computation, especially when referring to the heavy table interpolation processes in the QED-MC algorithm, since locating the random number $\xi$ in the table requires searching the array of the integral of $\frac{d^2N_\gamma}{d\tau d\chi_\gamma}$ for a given $\chi_e$, the speed of which is at the order of $O(N)$ or $O(\log N)$ when using bianry search.

By implementing the sigmoid sampling method into SMILEI, we perform simulations to compare the photon spectrums in Fig. 9 where the parameter of Fig. 9(b) is identical to Fig. 4. It can be seen that with Sigmoid modification one gets relatively more accurate results in the $\chi_e > 1$ region with table size of $N = 128$ when compared to the original sampling method with larger table sizes. In other words, by using Sigmoid like tabulation we can achieve more accurate photon emission in both classical and deep QED region, which significantly improve the applicability of PIC-MC simulations in the strong-field QED regime.

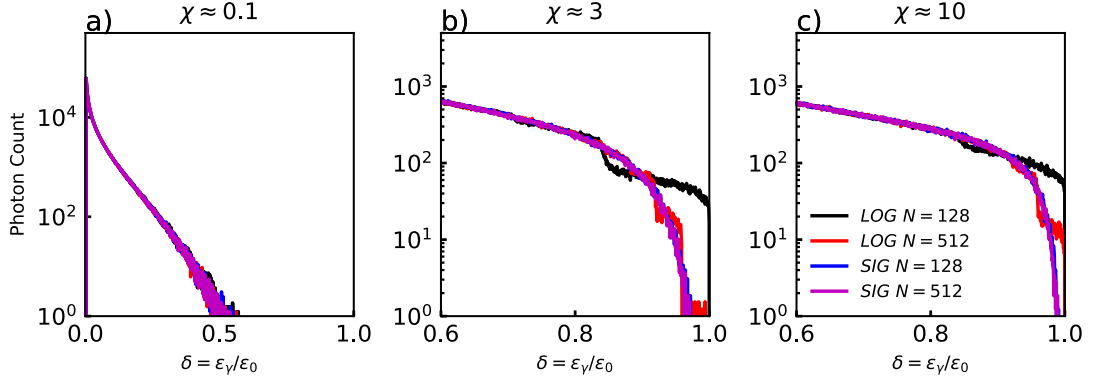

**Figure 9.** Photon energy spectrums in PIC simulations with $a_0 = 20, \tau_L = 5\text{fs}$. The electron energies are (a) $\varepsilon_e = 500\text{MeV}$, (b) $\varepsilon_e = 500\text{MeV}$ and (c) $\varepsilon_e = 5000\text{MeV}$, corresponding to $\chi_e \approx 0.1, 3, 10$. The results of Sigmoid sampling are shown by the blue and purple lines for table sizes of $N = 128$ and $N = 512$; logarithmic sampling results are shown by the black and red lines for $N = 128$ and $N = 512$.

**4.4 Connections to other numerical problems**

The numerical implementation of the Monte-Carlo QED process in PIC simulation has been discussed and tested in multiple research [14,17,19,26,27,31]. It is pointed out that common numerical implementation fails out of the valid region of LCFA, i.e., where the harmonic structure of the radiation spectrum and quantum interference effect are important. Therefore, recent studies focus on extending the valid region of LCFA [32–35]. In our work we find that insufficient sampling of the formulars based on LCFA leads to the numerical errors. Therefore, our work focuses on removing the numerical errors of these implementations and recovering the physical process that is improperly programed. On the other hand, the region where the numerical errors are significant is well within the valid region of LCFA where the field strength is sufficiently high, and the harmonic radiation and quantum interference can be neglected.

**5. Conclusion**

In conclusion, we have studied in detail the interaction between an ultrashort high intensity laser pulse and an ultra-relativistic electron beam. By using PIC simulation, we find the emergence of stair-like structure in the high energy part of the spectra in moderate quantum region. The structure is more apparent for higher electron QED parameters. We find out that the structure stems from inaccurate photon emission due to insufficient resolution of the tables used in the QED-MC algorithm. These may induce significant error for high energy photons and sequential electron-positron pair production. We therefore proposed a novel way using Sigmoid sampling to overcome the challenge, which effectively extends the application of the QED-MC algorithm in more extreme situations including high-energy photon emission, Breit-Wheeler pair-production by these photons and consequent QED cascades in the $\chi_e > 1$ region.

**Acknowledgement**

The authors would like to thank the contributors of the PIC code SMILEI [12]. This work was supported by the Strategic Priority Research Program of the Chinese Academy of Sciences (Grant No. XDB 16010000) and the National Natural Science Foundation of China (Grants No. 11875307, No.11935008 and No. 11804348).

**Declaration of competing interest**

The authors declare that they have no known competing financial interests or personal relationships that could have appeared to influence the work reported in this paper.

**Reference**

[1] J. W. Yoon, Y. G. Kim, I. W. Choi, J. H. Sung, H. W. Lee, S. K. Lee, and C. H. Nam, *Realization of Laser Intensity over $10^{23}$ W/Cm$^2$*, Optica **8**, 630 (2021).

[2] C. N. Danson, C. Haefner, J. Bromage, T. Butcher, J.-C. F. Chanteloup, E. A. Chowdhury, A. Galvanauskas, L. A. Gizzi, J. Hein, D. I. Hillier, N. W. Hopps, Y. Kato, E. A. Khazanov, R. Kodama, G. Korn, R. Li, Y. Li, J. Limpert, J. Ma, C. H. Nam, D. Neely, D. Papadopoulos, R. R. Penman, L. Qian, J. J. Rocca, A. A. Shaykin, C. W. Siders, C. Spindloe,

S. Szatmári, R. M. G. M. Trines, J. Zhu, P. Zhu, and J. D. Zuegel, *Petawatt and Exawatt Class Lasers Worldwide*, High Power Laser Sci. Eng. **7**, e54 (2019).

[3] Z. Zhang, F. Wu, J. Hu, X. Yang, J. Gui, P. Ji, X. Liu, C. Wang, Y. Liu, X. Lu, Y. Xu, Y. Leng, R. Li, and Z. Xu, *The Laser Beamline in SULF Facility*, High Power Laser Sci. Eng. **8**, (2020).

[4] T. G. Blackburn, *Radiation Reaction in Electron–Beam Interactions with High-Intensity Lasers*, Rev. Mod. Plasma Phys. **4**, 5 (2020).

[5] J. Dawson, *One-Dimensional Plasma Model*, Phys. Fluids **5**, 445 (1962).

[6] C. P. Ridgers, T. G. Blackburn, D. D. Sorbo, L. E. Bradley, C. Slade-Lowther, C. D. Baird, S. P. D. Mangles, P. McKenna, M. Marklund, C. D. Murphy, and A. G. R. Thomas, *Signatures of Quantum Effects on Radiation Reaction in Laser–Electron-Beam Collisions*, J. Plasma Phys. **83**, (2017).

[7] T. G. Blackburn, C. P. Ridgers, J. G. Kirk, and A. R. Bell, *Quantum Radiation Reaction in Laser–Electron-Beam Collisions*, Phys. Rev. Lett. **112**, 015001 (2014).

[8] C. N. Harvey, A. Gonoskov, A. Ilderton, and M. Marklund, *Quantum Quenching of Radiation Losses in Short Laser Pulses*, Phys. Rev. Lett. **118**, 105004 (2017).

[9] A. Hützen, J. Thomas, J. Böker, R. Engels, R. Gebel, A. Lehrach, A. Pukhov, T. P. Rakitzis, D. Sofikitis, and M. Büscher, *Polarized Proton Beams from Laser-Induced Plasmas*, High Power Laser Sci. Eng. **7**, (2019).

[10] M. Büscher, A. Hützen, L. Ji, and A. Lehrach, *Generation of Polarized Particle Beams at Relativistic Laser Intensities*, High Power Laser Sci. Eng. **8**, e36 (2020).

[11] T. D. Arber, K. Bennett, C. S. Brady, A. Lawrence-Douglas, M. G. Ramsay, N. J. Sircombe, P. Gillies, R. G. Evans, H. Schmitz, A. R. Bell, and C. P. Ridgers, *Contemporary Particle-in-Cell Approach to Laser-Plasma Modelling*, Plasma Phys. Control. Fusion **57**, 113001 (2015).

[12] J. Derouillat, A. Beck, F. Pérez, T. Vinci, M. Chiaramello, A. Grassi, M. Flé, G. Bouchard, I. Plotnikov, N. Aunai, J. Dargent, C. Riconda, and M. Grech, *Smilei : A Collaborative, Open-Source, Multi-Purpose Particle-in-Cell Code for Plasma Simulation*, Comput. Phys. Commun. **222**, 351 (2018).

[13] I. V. Sokolov, N. M. Naumova, and J. A. Nees, *Numerical Modeling of Radiation-Dominated and Quantum-Electrodynamically Strong Regimes of Laser-Plasma Interaction*, Phys. Plasmas **18**, 093109 (2011).

[14] A. Gonoskov, S. Bastrakov, E. Efimenko, A. Ilderton, M. Marklund, I. Meyerov, A. Muraviev, A. Sergeev, I. Surmin, and E. Wallin, *Extended Particle-in-Cell Schemes for Physics in Ultrastrong Laser Fields: Review and Developments*, Phys. Rev. E **92**, 023305 (2015).

[15] N. V. Elkina, A. M. Fedotov, I. Yu. Kostyukov, M. V. Legkov, N. B. Narozhny, E. N. Nerush, and H. Ruhl, *QED Cascades Induced by Circularly Polarized Laser Fields*, Phys. Rev. Spec. Top. - Accel. Beams **14**, 054401 (2011).

[16] A. Pukhov, *Three-Dimensional Electromagnetic Relativistic Particle-in-Cell Code VLPL (Virtual Laser Plasma Lab)*, J. Plasma Phys. **61**, 425 (1999).

[17] C. P. Ridgers, J. G. Kirk, R. Duclous, T. G. Blackburn, C. S. Brady, K. Bennett, T. D. Arber, and A. R. Bell, *Modelling Gamma-Ray Photon Emission and Pair Production in High-Intensity Laser–Matter Interactions*, J. Comput. Phys. **260**, 273 (2014).

[18]  M. Lobet, E. d'Humières, M. Grech, C. Ruyer, X. Davoine, and L. Gremillet, *Modeling of Radiative and Quantum Electrodynamics Effects in PIC Simulations of Ultra-Relativistic Laser-Plasma Interaction*, J. Phys. Conf. Ser. **688**, 012058 (2016).

[19]  R. Duclous, J. G. Kirk, and A. R. Bell, *Monte Carlo Calculations of Pair Production in High-Intensity Laser–Plasma Interactions*, Plasma Phys. Control. Fusion **53**, 015009 (2011).

[20]  *Generation of the External Tables — Smilei 4.6 Documentation*, https://smileipic.github.io/Smilei/tables.html.

[21]  J. Schwinger, *On Gauge Invariance and Vacuum Polarization*, Phys. Rev. **82**, 664 (1951).

[22]  V. I. Ritus, *Quantum Effects of the Interaction of Elementary Particles with an Intense Electromagnetic Field*, J. Sov. Laser Res. **6**, 497 (1985).

[23]  A. D. Piazza, *Extremely High-Intensity Laser Interactions with Fundamental Quantum Systems*, Rev Mod Phys **84**, 52 (2012).

[24]  A. G. R. Thomas, C. P. Ridgers, S. S. Bulanov, B. J. Griffin, and S. P. D. Mangles, *Strong Radiation-Damping Effects in a Gamma-Ray Source Generated by the Interaction of a High-Intensity Laser with a Wakefield-Accelerated Electron Beam*, Phys. Rev. X **2**, 041004 (2012).

[25]  A. Nikishov and V. Ritus, *Quantum Processes in the Field of a Plane Electromagnetic Wave and in a Constant Field .1.*, Sov. Phys. Jetp-Ussr **19**, 529 (1964).

[26]  C. N. Harvey, A. Ilderton, and B. King, *Testing Numerical Implementations of Strong-Field Electrodynamics*, Phys. Rev. A **91**, 013822 (2015).

[27]  T. G. Blackburn, D. Seipt, S. S. Bulanov, and M. Marklund, *Benchmarking Semiclassical Approaches to Strong-Field QED: Nonlinear Compton Scattering in Intense Laser Pulses*, Phys. Plasmas **25**, 083108 (2018).

[28]  J. Esberg and U. I. Uggerhøj, *Does Experiment Show That Beamstrahlung Theory - Strong FIeld QED - Can Be Trusted?*, J. Phys. 16 (2009).

[29]  L. L. Ji, J. Snyder, and B. F. Shen, *Single-Pulse Laser-Electron Collision within a Micro-Channel Plasma Target*, Plasma Phys. Control. Fusion **61**, 065019 (2019).

[30]  T. G. Blackburn, A. Ilderton, M. Marklund, and C. P. Ridgers, *Reaching Supercritical Field Strengths with Intense Lasers*, New J. Phys. **21**, 053040 (2019).

[31]  T. G. Blackburn, A. J. MacLeod, and B. King, *From Local to Nonlocal: Higher Fidelity Simulations of Photon Emission in Intense Laser Pulses*, New J. Phys. **23**, 085008 (2021).

[32]  A. Di Piazza, M. Tamburini, S. Meuren, and C. H. Keitel, *Implementing Nonlinear Compton Scattering beyond the Local-Constant-Field Approximation*, Phys. Rev. A **98**, 012134 (2018).

[33]  A. Di Piazza, M. Tamburini, S. Meuren, and C. H. Keitel, *Improved Local-Constant-Field Approximation for Strong-Field QED Codes*, Phys. Rev. A **99**, 022125 (2019).

[34]  A. Ilderton, B. King, and D. Seipt, *Extended Locally Constant Field Approximation for Nonlinear Compton Scattering*, Phys. Rev. A **99**, 042121 (2019).

[35]  T. Heinzl, B. King, and A. J. MacLeod, *Locally Monochromatic Approximation to QED in Intense Laser Fields*, Phys. Rev. A **102**, 063110 (2020).